\documentclass[aps,prl,twocolumn,floatfix,citeautoscript,nofootinbib,superscriptaddress,10pt]{revtex4-2}

\usepackage{orcidlink}
\usepackage[T1]{fontenc}
\usepackage{graphicx}
\usepackage[sectionbib]{bibunits}
\usepackage{bibunits}
\usepackage{dcolumn}
\usepackage{pifont}
\usepackage{bm}
\usepackage{braket}
\usepackage{amsmath}
\usepackage{mathtools}
\usepackage{graphicx,color,xcolor}
\usepackage{hyperref}
\usepackage[utf8]{inputenc}
\usepackage[capitalize]{cleveref}
\newcommand{\abs}[1]{\left| #1 \right|} 

\usepackage{multirow}

\usepackage{soul} 

\usepackage[normalem]{ulem} 

\begin{document}

\begin{bibunit}[apsrev4-2]

\title{First-order phase transition in atom-molecule quantum degenerate mixtures with coherent three-body recombination}

\author{G. A. Bougas \orcidlink{0009-0002-5643-9293}}
\affiliation{Department of Physics and LAMOR, Missouri University of Science and Technology, Rolla, MO 65409, USA}

\author{A. Vardi \orcidlink{0000-0002-8992-2129} }
\affiliation{Department of Chemistry, Ben-Gurion University of the Negev, Beer-Sheva 84105, Israel}
\affiliation{ITAMP, Center for Astrophysics $|$ Harvard \& Smithsonian Cambridge, Massachusetts 02138, USA}

\author{H.~R.~Sadeghpour\orcidlink{0000-0001-5707-8675}}
\affiliation{ITAMP, Center for Astrophysics $|$ Harvard \& Smithsonian Cambridge, Massachusetts 02138, USA}

\author{C. Chin \orcidlink{0000-0003-0278-4630} }
\affiliation{James Franck Institute, Enrico Fermi Institute and Department of Physics,
University of Chicago, Chicago, Illinois 60637, USA}

\author{S. I. Mistakidis \orcidlink{0000-0002-5118-5792}}
\affiliation{Department of Physics and LAMOR, Missouri University of Science and Technology, Rolla, MO 65409, USA}

\date{\today}

\begin{abstract}

We map the phase diagram of a two-mode atom-molecule Bose-Einstein condensate with Fano-Feshbach and coherent three-body recombination (cTBR) terms. The standard second order phase transition observed as the molecular energy is tuned through 
the Fano-Feshbach resonance, is replaced by a first  order transition when cTBR becomes prominent, due to a double-well structure in the free energy landscape. This transition is associated with atom-molecule entanglement, bistability, and molecular metastability. Our results establish cTBR as a powerful knob for quantum state engineering and control of reaction dynamics in ultracold chemistry.

\end{abstract}

\maketitle

Ultracold molecular Bose-Einstein condensates (mBECs) represent an emergent platform for precision measurements~\cite{Safronova_review,roussy2023improved,Arrowsmith-Kron_2024,demille2024quantum}, quantum simulation~\cite{Capogrosso_polar,Gorshkov_tunable,Pollet_SS,Hazzard_magnetism,cornish2024quantum}, and  information~\cite{asnaashari2023general,sawant2020ultracold,Wang_Rydberg,Hughes_entanglement}, as well as quantum chemistry~\cite{bohn2017cold,rui2017controlled,Wolf_state_2017,hoffmann2018reaction,martins2025microwave}. 
They offer unprecedented tunability in terms of the involved microscopic degrees-of-freedom~\cite{burchesky2021rotational,softley_cold_2023,langen2024quantum}, which promises a wide range of applications in the broader field of quantum science and technology~\cite{carr2009cold,softley_cold_2023}. 
Experimentally available settings consist of either polar molecules such as KRb~\cite{de2019degenerate}, NaK~\cite{wu2012ultracold}, RbCs~\cite{molony2014creation} and NaCs~\cite{bigagli2024observation} or weakly bound diatomic Fano-Feshbach molecules of Rb$_{2}$~\cite{haze2025controlling,Wolf_state_2017,dorer2025steering} and Cs$_{2}$~\cite{Zhang_transition_2021,zhang2023many}. 

Fano-Feshbach molecules~\cite{donley2002atom,Durr,thalhammer2006long,Cheng_sensitive_2003}, on which we focus herein, are commonly produced via magnetoassociation~\cite{kohler2006production} converting pairs of atoms into diatomic molecules through interaction ramps in the vicinity of a closed-channel (typically narrow) $g$-wave resonance~\cite{zhang2023many} supporting collisional stability~\cite{chin2010feshbach,mark2018mott}. 
Recent experimental observations revealed the phenomenon of collective, Bose-enhanced superchemistry~\cite{zhang2023many} (see also the original  theory works~\cite{heinzen2000superchemistry,moore2002bose}), ensuing quantum phase transitions~\cite{Zhang_transition_2021,zhang2023many}, and accompanied entanglement processes~\cite{nagata2025phase} as well as opportunities for controlling chemical reaction pathways emulating beam splitter analogues~\cite{dorer2025steering}.  

\begin{figure}[t!]
\centering
\includegraphics[width=1\columnwidth]{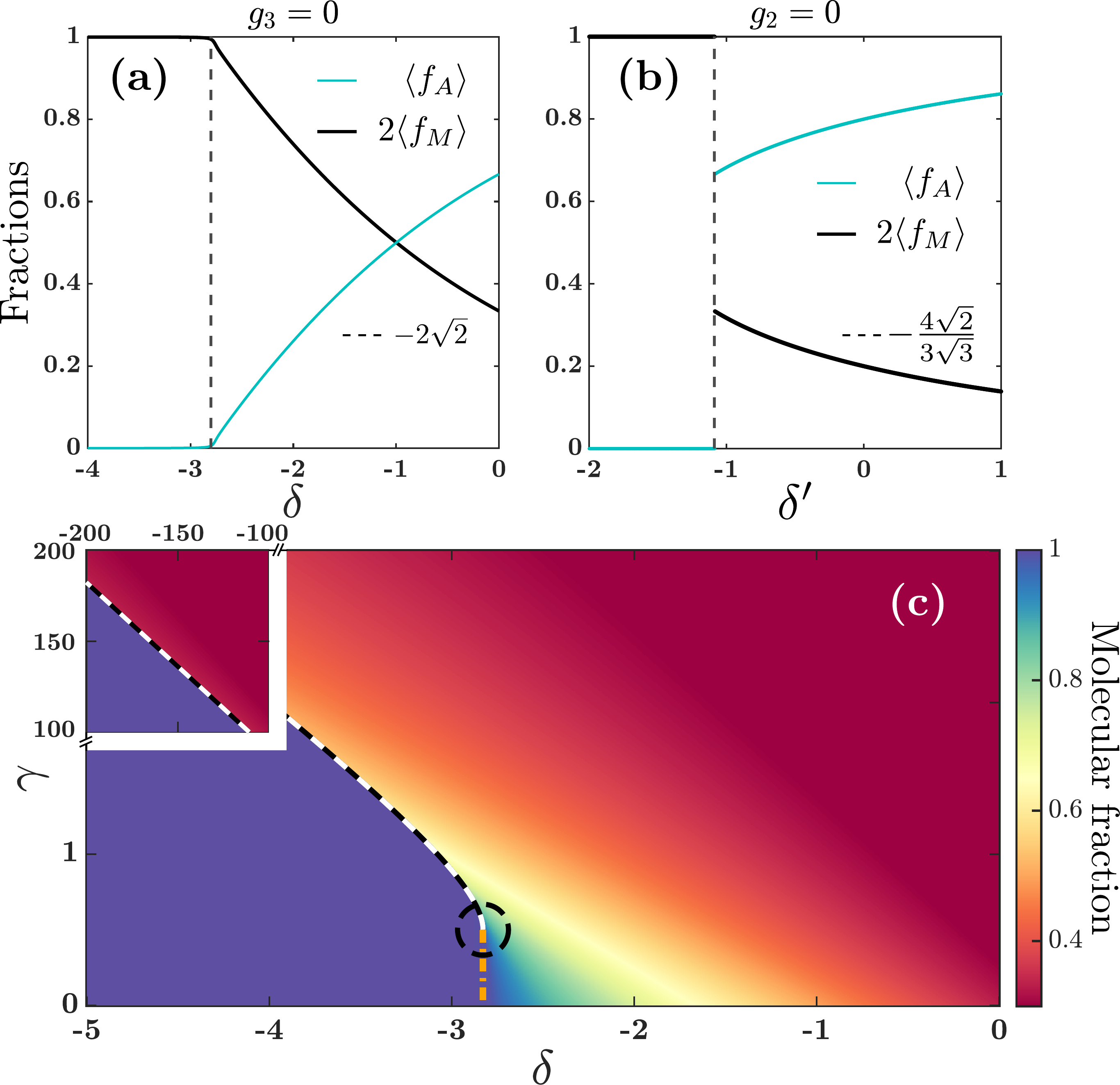}
\caption{First- and second-order phase transitions of atom-molecules with competing Fano-Feshbach coupling and cTBR. (a) In the case of pure Fano-Feshbach coupling, a second-order transition takes place (dashed line), from a pure molecular state to a mixture of atoms (bright teal line) and molecules (black line). (b) When only three-body coupling is present, the transition becomes first-order, denoted by the vertical dashed line. (c) Molecular fraction, $2\braket{f_M}$, over the $(\delta,\gamma)$ plane. The black-white dashed (orange dash-dotted) curve delineates the first-order (second-order) transition, while the dashed circle marks the tricritical point at their intersection. The box in the upper left corner presents the regime where cTBR prevails.
The particle number is $N=800$.}
\label{Fig:Phase_diagram}
\end{figure}

At quantum degeneracy, the reaction dynamics stems from the nonlinear matter-wave mixing between reactants and products dictated by the following 
processes: i) reversible Fano-Feshbach coupling~\cite{Falco} associating two atoms into a weakly bound diatomic molecule  
and ii) irreversible three-body recombination~\cite{haze2025controlling,Stevenson} (TBR) related to three atom collision yielding a diatomic molecule and a free atom.
The latter process is also encountered in dissociative recombination mechanisms occurring in diffuse interstellar clouds~\cite{Kokoouline_mechanisms_2001,McCall_detection_1998}.

In trapped ultracold gases, TBR 
is a loss mechanism \cite{weber2003three,fletcher2013stability,Greene_review}, since the resulting molecule and the remaining free atom acquire large kinetic energies due to energy conservation,
and thus escape from the trap; the zero-energy three-atom collision dictates that the kinetic energies of both products counterbalance the molecular binding energy.
TBR limits
the production of quantum-degenerate gases as the gas parameter $\sqrt{a_s^3 n}$ (with $a_s$ and $n$ being the $s$-wave scattering length and the atom density, respectively) approaches unitarity, or interfering with their conversion via magnetic Fano-Feshbach-resonances~\cite{chin2010feshbach} or optical  photoassociation~\cite{Jones_review}, into mBECs. However, recent experiments~\cite{zhang2023many} have demonstrated the possibility of {\em coherent} TBR (cTBR), 
in which three atoms are converted to a weakly bound Fano-Feshbach molecule and a free atom in a reversible way. Owing to the small deep-dimer relaxation rates~\cite{Regal_lifetime_2004} in the vicinity of a Fano-Feshbach resonance, such a process can be considered as nearly elastic. Signatures of cTBR have been observed in the atom-molecule population  oscillations, in particular in the dependence of their frequency on the particle number~\cite{zhang2023many}.
In spite of incoherent losses, a saturation trend was observed in the relevant population fractions, motivating the investigation of the impact of cTBR on  the atom-molecule phase diagram and the associated dynamics.

\begin{figure}[t!]
\centering
\includegraphics[width=1\columnwidth]{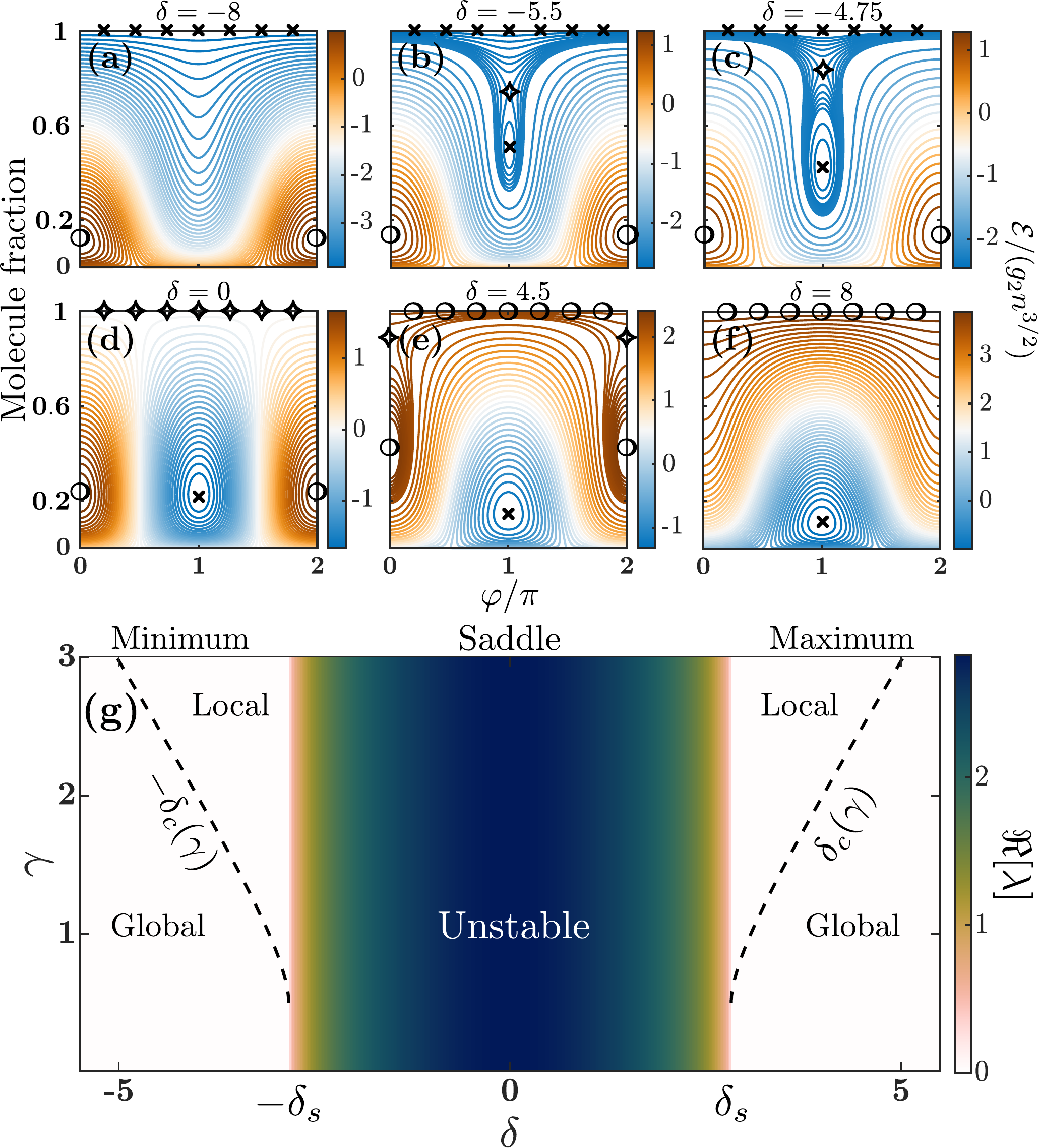}
\caption{
Phase-space portraits and stability diagram of the atom-molecule phases. 
(a)-(f) Energy density isolines for $\gamma = 3$ and various $\delta$ (see legends) on the plane defined by the molecular fraction, $2f_M$, and the relative angle between atoms and molecules, $\varphi$. The x, \ding{109}, \ding{71} symbols mark the energy minima, maxima and saddle points respectively. 
The $f_M=0.5$ line (mBEC) refers to a single fixed point for which $\varphi$ is ill-defined, while the energy maxima (same fixed point) at $\varphi=0,2\pi$ corresponds to full circles about the $2f_M$ axis.
(g) 
Stability diagram of the molecular condensate.
The presence of a real part of the eigenvalues, $\lambda$, pertaining to the linearized Hamiltonian system around $f_M=1/2$ leads to instability within $\abs{\delta} \leq \delta_s$. In the stable regions, the mBEC can be either a global or a local minimum or maximum, due to the double-well.}
\label{Fig:Energy_contours}
\end{figure}

Here, we study the ground state phase diagram of two-mode atom-molecule condensates in the presence of both Fano-Feshbach and cTBR couplings, see Fig.~\ref{Fig:Phase_diagram}. When cTBR dominates over the direct Fano-Feshbach coupling,  we find a striking first-order phase-transition in which the molecular ground state occupation drops discontinuously as the atom-molecule detuning is varied. This first-order transition should be contrasted with the familiar second-order transition when the direct coupling is dominant~\cite{Romans_quantum_2004,Radzihovsky_transition,Zhang_transition_2021}. The transition originates from the appearance of a double-well structure [see Fig.~\ref{Fig:Energy_contours}(b),(c)] in the mean-field free energy of the atom-molecule system, with one minimum being the molecular condensate and the other being a mostly-atoms coherent superposition. 
In the vicinity of the transition, the system is bistable, resulting in a highly nonclassical ground state exhibiting large atom-molecule entanglement with a tendency to approach an atom-molecule cat state. We propose quench protocols to dynamically demonstrate the metastability of the molecular condensate past the phase transition, as opposed to the molecular instability in the case of pure Fano-Feshbach coupling~\cite{Barankov_coexistence,Basu_stability,Wang_tunable_2025}, and highlight its dependence on the atom number.

\paragraph*{\textit {Phase transitions.}}
We consider an ensemble of $N_A$ atoms and $N_M$ molecules, with relevant fractions $f_A = N_A/N$, $f_M=N_M/N$, satisfying $1=f_A+2f_M$, within a box of volume $V$
and density $n=N/V$.
The minimal model for capturing atom-molecule coherence induced by both the Fano-Feshbach coupling and cTBR is the two-mode Hamiltonian~\cite{vardi2001quantum,Kostrun_theory_2000} (for more details see Supplementary Material (SM)~\cite{supp}),
\begin{equation}
\hat{H} = \Delta \hat{b}^{\dagger} \hat{b} + \frac{g_2}{V^{1/2}} \left(  \hat{b}^{\dagger} \hat{a}^2 + \text{h.c.}  \right) + \frac{g_3}{V^{3/2}} \left(  \hat{b}^{\dagger} \hat{a}^{\dagger} \hat{a}^3  + \text{h.c.}  \right).
\label{Eq:Hamiltonian}
\end{equation}
Here $\hat{a}$ ($\hat{b}$) is the bosonic operator annihilating one atom (molecule), while $\text{h.c.}$ stands for the hermitian conjugate.
The detuning $\Delta$ determines the dressed energy of the Fano-Feshbach molecule, and it is experimentally tunable by external magnetic fields~\cite{chin2010feshbach,Zhang_transition_2021}.
The second term in Eq.~\eqref{Eq:Hamiltonian} refers to the Fano-Feshbach mechanism (converting two atoms into a molecule and vice versa) 
with coupling $g_2$~\cite{Duine_atom_2004}. 
The last term in Eq.~\eqref{Eq:Hamiltonian} accounts for the reversible cTBR (where three atoms collide forming a molecule and a free atom) of strength $g_3$. 
To emphasize the universal features of the emergent phase transitions, we define the characteristic parameters of the two-mode model, $\delta = \Delta/(g_2 \sqrt{n})$, and $\gamma = g_3 n / g_2$. These dimensionless parameters determine the nature of the two-mode dynamics and will later appear as tunable parameters in the mean-field energy density [see Eq.~\eqref{Eq:Energy_combined}]. 
To make a connection with experimental observations, the detuning $\Delta$ is expressed in units of the Fano-Feshbach resonance width~\cite{Wang_tunable_2025}, $\Delta\mu \Delta B$, by multiplying $\delta$ with the dimensionless ratio, $\left(  \frac{2\pi \hbar^2 n a_{bg}}{m \Delta \mu \Delta B}  \right)^{1/2}$.
Here, $a_{bg}$ is the background atom-atom scattering length, $\hbar$ is the reduced Planck constant, $m$ is the atom mass, $\Delta B$ is the width of the resonance, while $\Delta \mu$ is the difference between the magnetic moments of two atoms and a molecule. 
Many-body atom-molecule eigenstates are obtained by exact diagonalization of 
the Hamiltonian in Eq.~(\ref{Eq:Hamiltonian}),  represented in the Fock basis $\{ \ket{n_M} \equiv \ket{n_M; N-2n_M}, ~ n_M=0,\ldots,N/2\}$.

The first step in understanding the atom-molecule coexistence is to study their ground state phase diagram~\cite{Radzihovsky_transition,Romans_quantum_2004,Wang_tunable_2025}.
Considering vanishing cTBR (i.e., $g_3=0$), a well known second-order phase transition occurs at the critical detuning $\delta=-\delta_s=-2\sqrt{2}$~\cite{Tikhonenkov_many_2006,Pazy_nonlinear_2005}, see vertical dashed line in Fig.~\ref{Fig:Phase_diagram}(a). 
In the case of the narrow resonance of $19.8~\rm{G}$ in $^{133}$Cs and a density $n=2.9\times10^{13}~ \rm{cm}^{-3}$~\cite{Wang_stability_2024}, $\delta_s$ corresponds to $\Delta=-0.38 ~ \Delta\mu \Delta_B$. 
For $\delta \leq -\delta_s$, all $N=800$ atoms (bright teal solid line) are converted into molecules (black solid line).
However, for $\delta > -\delta_s$ the molecular fraction, $2\braket{f_M}$, decreases continuously and asymptotically approaches zero as $\delta \to + \infty$. In sharp contrast, the phase transition to a molecular condensate becomes first-order when solely considering cTBR ($g_2=0$). 
This transition takes place at $\delta'=-\delta'_c  = - 4\sqrt{2}/(3\sqrt{3})$, where $\delta'=\Delta/(g_3 n^{3/2})$ [vertical dashed line in Fig.~\ref{Fig:Phase_diagram}(b)].
Specifically, for $\delta' > -\delta'_c$, the molecular fraction abruptly jumps to $2\braket{f_M} \simeq 1/3$ and then steadily decreases.

The feasibility of experimental observation of the first-order phase transition hinges on the values of the detuning, $\delta$, and coupling ratio, $\gamma$, at which the phase transition is modified. 
For extremely large $\gamma$, where $g_3 n \gg g_2$, the relevant phase diagram with combined reaction mechanisms depicted in Fig.~\ref{Fig:Phase_diagram}(c) reveals the existence of a first-order transition for large negative $\delta$ [top-left box in Fig.~\ref{Fig:Phase_diagram}(c)].
The $\gamma$ dependent transition boundary is located along $\delta' \equiv \delta /\gamma=-\delta'_c$ [black-white dashed line in the top-left box of Fig.~\ref{Fig:Phase_diagram}(c) and in Fig.~\ref{Fig:Phase_diagram}(b)]. Since $\gamma$ scales with $n$, this means that the critical detuning for the first-order phase transition in the presence of pure cTBR coupling is vastly larger than that of the Fano-Feshbach-coupling second-order transition, rendering its observation seemingly unrealistic. However, the phase transition retains its  first-order nature even at much smaller coupling ratios, e.g. $\gamma = 1$. 
Hence, it should be experimentally accessible owing to the moderate values of $\gamma$ required.
In particular, in the recent ${}^{133}$Cs experiment~\cite{zhang2023many} the reported $g_3$ refers to $\gamma \simeq 0.7$. 
Furthermore, the order of the phase transitions is maintained even in the presence of quantum fluctuations, intraspecies interactions and kinetic terms, see also SM~\cite{supp}.

\paragraph*{\textit {Mean-field energy density}.}

The nature of the phase-transitions in Fig.~\ref{Fig:Phase_diagram} is readily explained in terms of the mean-field energy density (see also SM~\cite{supp}) of the combined-coupling Hamiltonian of  Eq.~(\ref{Eq:Hamiltonian}), 
\begin{equation}
\frac{\mathcal{E}}{g_2 n^{3/2}} = \delta f_M  
 + 2 \sqrt{f_M}(1-2f_M) \left[ 1  + \gamma (1-2f_M)    \right] \cos(\varphi).
\label{Eq:Energy_combined}
\end{equation}
Here, 
$\varphi=2\theta_A-\theta_M$ is the relative phase between atoms ($\theta_A$) and molecules ($\theta_M$)~\cite{nagata2025phase}. Representative plots of this classical potential surface at $\gamma=3$ and various values of $\delta$, are shown in Fig.~\ref{Fig:Energy_contours}(a)-(f), with the fixed points satisfying $\dot{f}_M=\dot{\varphi}=0$, marked by $\times$, \ding{109}, \ding{71} for minima, maxima, and saddle points, respectively.

\begin{figure}[t!]
\centering
\includegraphics[width=1\columnwidth]{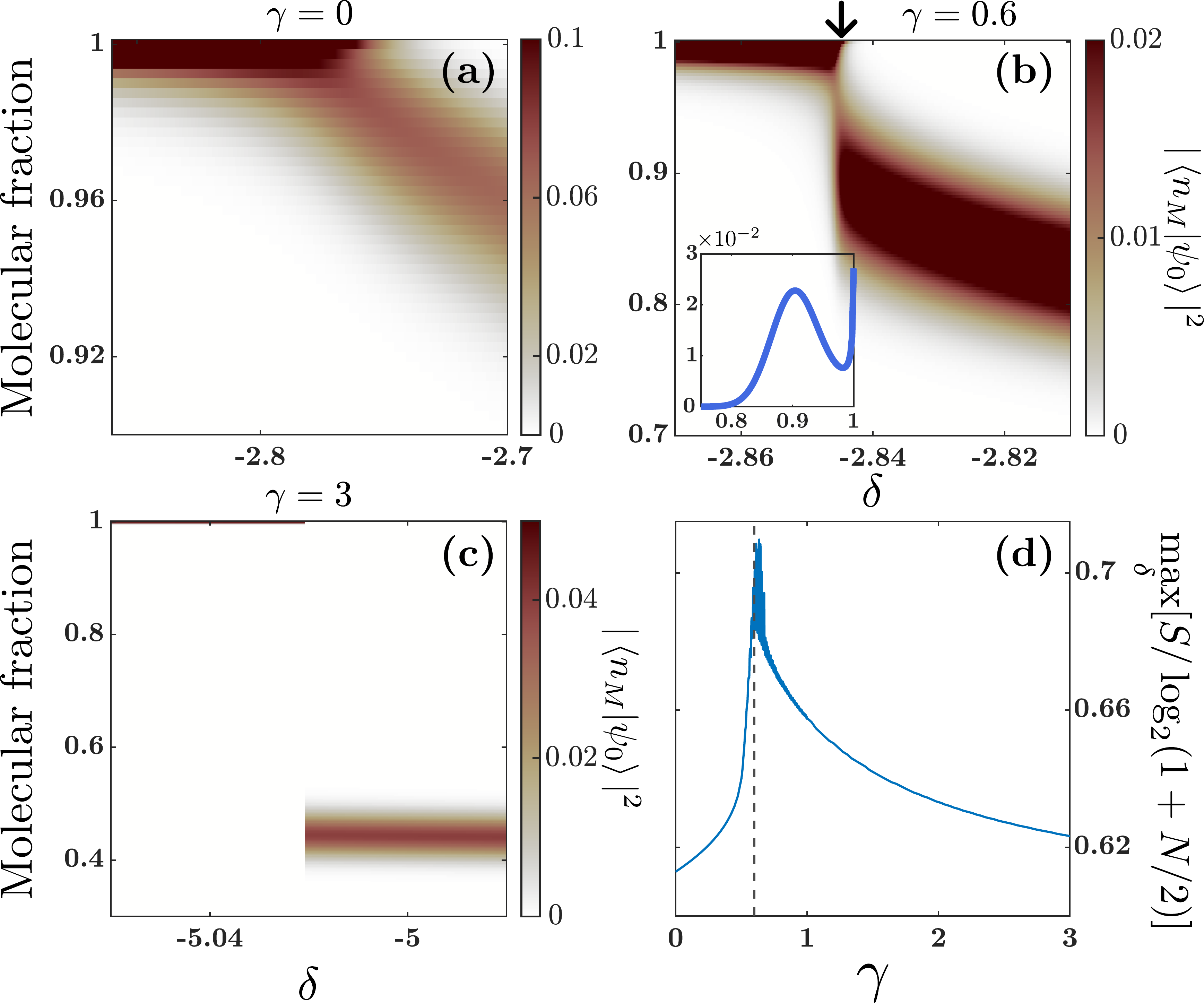}
\caption{
(a)-(c) Distribution of number states, $p_n \equiv\abs{\braket{n_M | \psi_0}}^2$, close to the transition point for different $\gamma$ (see legends). A superposition of a pure molecular condensate and a finite atom-molecule mixture appears at $\gamma = 0.6$ [panel (b)], due to the double-well potential in the mean-field energy [Fig.~\ref{Fig:Energy_contours}(c)]. The inset in panel (b) presents a distribution profile at the position indicated by the arrow. 
(d) Maximum of the normalized von Neumann entropy with respect to $\delta$. Its maximum value is attained at $\gamma = 0.6$ (dashed line), in the vicinity of the phase transition character change (from second- to first-order).}
\label{Fig:Entropy}
\end{figure}

As seen in Fig.~\ref{Fig:Energy_contours}(a), at large negative detuning, the all-molecules state is a global minimum, whereas the maximum refers to a mostly atoms state. However, as the detuning reaches the value $\delta=-\delta_d(\gamma)$, where (see SM~\cite{supp}),
\begin{equation}
\delta_d(\gamma) = \frac {  \left(12\gamma+6 \right) \sqrt{S}  - 16\sqrt {3}  \left(\gamma^2 +\gamma-\frac{3}{8} \right)  } {3\sqrt {5} \sqrt{ \gamma \sqrt { 3S } + 3\gamma + 6 \gamma^2}}, 
\label{Eq:Potential_existence}
\end{equation}
$S=32\gamma(\gamma+1)+3$, the ground state bifurcates into a double-well structure, consisting of two minima separated by a saddle point [Fig.~\ref{Fig:Energy_contours}(b)]. One of these two minima is the molecular BEC, whereas the other one has an atomic component that grows as the detuning is further increased [Fig.~\ref{Fig:Energy_contours}(c)]. At $\delta=-\delta_c(\gamma)$, where (see SM~\cite{supp}),
\begin{equation}
\delta_c(\gamma) = \frac{4\sqrt{2}}{3\sqrt{3}} 
\frac{(1+\gamma)^{3/2}}{\sqrt{\gamma}},
\end{equation}
this new minimum's energy becomes smaller than that of the all-molecules state, resulting in the first-order phase transition. The finite molecular fraction in the new global minimum at the transition point reads  $2f^{(c)}_M=(1+\gamma)/(3\gamma)$. 

The mean-field potential analysis  is in excellent agreement with the many-body numerical diagonalization results (see $-\delta_c(\gamma)$ black-white dashed line in Fig.~\ref{Fig:Phase_diagram}(c)). 
In fact, for $N>1000$, the two-mode model results converge to the thermodynamic limit (mean-field) predictions, as can be deduced from a finite-size analysis of distinct observables.
In particular, it is easily verified that $\delta_c(\gamma)/\gamma\rightarrow \delta'_c$ and $2\braket{f^{(c)}_M}\rightarrow 1/3$ in the limit $\gamma\gg 1$, as noted above for pure cTBR coupling [Fig.~\ref{Fig:Phase_diagram}(b)].                                   
The double-well structure in Fig.~\ref{Fig:Energy_contours} only appears for $\gamma>0.5$ (see SM~\cite{supp}), and thus the transition is first-order within this interval. In the regime of $\gamma \in [0,0.5)$, the absence of a double-well structure in the energy density facilitates the second-order transition at $\delta_c(\gamma) = \delta_s$ [orange dash-dotted line in Fig.~\ref{Fig:Phase_diagram}~(c)], as for the pure Fano-Feshbach mechanism [Fig.~\ref{Fig:Phase_diagram}(a)]. 
The intersection of the two distinct phase transition boundaries hints at the appearance of a tricritical point~\cite{Griffiths_phase_1975,Chang_generalized_1973,Cheng_chiral_2022}, marked by the dashed circle in Fig.~\ref{Fig:Phase_diagram}(c), see also SM~\cite{supp} for the scaling of the molecular fraction close to the transition point.

\paragraph*{\textit {Molecular metastability.}}
Linearizing the Hamiltonian system [Eq.~\eqref{Eq:Energy_combined}] around the stationary points, their stability
is inferred from the absence of a real part in the eigenvalues, $\lambda$, of the relevant Jacobian [see also SM~\cite{supp}].
In particular, the mBEC is unstable (a hyperbolic saddle point) for $\abs{\delta} \leq \delta_s$, regardless of $\gamma$ [Fig.~\ref{Fig:Energy_contours}(g)]. Since $\delta_d(\gamma)>\delta_c(\gamma)>\delta_s$, we conclude that the introduction of cTBR results in {\em bistability} in the range $\delta\in[-\delta_d(\gamma),-\delta_s]$ [Fig.~\ref{Fig:Energy_contours}(g)], with the mBEC remaining metastable beyond the phase transition, i.e. in the range $\delta\in[-\delta_c(\gamma),-\delta_s]$. 
This is in stark contrast to the pure Fano-Feshbach coupling where the mBEC is an unstable saddle point beyond the phase transition.

The energy landscape at $\delta>0$  ([Fig.~\ref{Fig:Energy_contours}(d)-(f)]) reflects the $(\delta,\varphi,\mathcal{E}) \rightarrow (-\delta,\varphi+\pi,-\mathcal{E})$  
symmetry of the energy density, with molecular stability restored at $\delta=\delta_s$.
Accordingly, the double-well structure along $\varphi=\pi$ is replaced by a double-hump along $\varphi=0$ for $\delta\in[\delta_s,\delta_d(\gamma)]$ [Fig.~\ref{Fig:Energy_contours}(e)]. For $\delta>\delta_d(\gamma)$ the mBEC becomes a global maximum with the ground state being mostly atoms [Fig.~\ref{Fig:Energy_contours}(f)], mirroring Fig.~\ref{Fig:Energy_contours}(a).

\paragraph*{\textit{Atom-molecule entanglement.}} 
The entanglement entropy between atoms and molecules is $S=-{\rm Tr}_M\left(\hat{\rho}_A\log_2\hat{\rho}_A\right)=-{\rm Tr}_A\left(\hat{\rho}_M\log_2\hat{\rho}_M\right)$ where $\hat{\rho}_A$, ($\hat{\rho}_M$) is the reduced atomic (molecular) density matrix, obtained by a partial trace over molecular (atomic) degrees of freedom. For the two-mode model, where the numbers of atoms and molecules are interrelated, 
this definition coincides with the Shannon entropy ${\cal H}=-\sum_n p_n\log_2 p_n$,  where $p_n\equiv\abs{\braket{n_M | \psi_0}}^2$ is the ground state's probability distribution in the number state (a.k.a `computational') basis. The degree of atom-molecule entanglement is thus directly related to the Fock state participation (number of contributing states) of the ground state.

The $p_n$ distribution is presented as a function of $\delta$ for different values of $\gamma$ in Fig.~\ref{Fig:Entropy}(a)-(c). For $\delta<-\delta_c(\gamma)$, the quantum ground state is supported by the all-molecules well, resulting in a low-participation narrow distribution about the $|n_{M}=N/2\rangle$ state. In all cases, $p_n$ broadens past the transition point and disperses around a $\gamma$-dependent finite atom-molecule fraction, which in the presence of the double-well potential [Fig.~\ref{Fig:Entropy}(b), (c)] is approximately $(1+\gamma)/(3\gamma)$.

Focusing on Fig.~\ref{Fig:Entropy}(b), we note that the $p_n$ width in the vicinity of the critical detuning is particularly large, due to the delocalization of the quantum state between the two wells. The resulting broad bimodal probability distribution [inset of Fig.~\ref{Fig:Entropy}(b)] approaches as $N$ is increased, a macroscopic cat state~\cite{Yurke_generating_1986,Friedman_quantum_2000,Arndt_Testing_2014,Qin_generating_2021}. As discussed above, the $p_n$ broadening is associated with high atom-molecule entanglement entropy [see Fig.~\ref{Fig:Entropy}(d)]. We thus conclude that cTBR leads to much stronger atom-molecule entanglement in the vicinity of the first-order phase transition, than that obtained using pure Fano-Feshbach coupling \cite{Hines_entanglement_2003}.

\begin{figure}[t!]
\centering
\includegraphics[width=1\columnwidth]{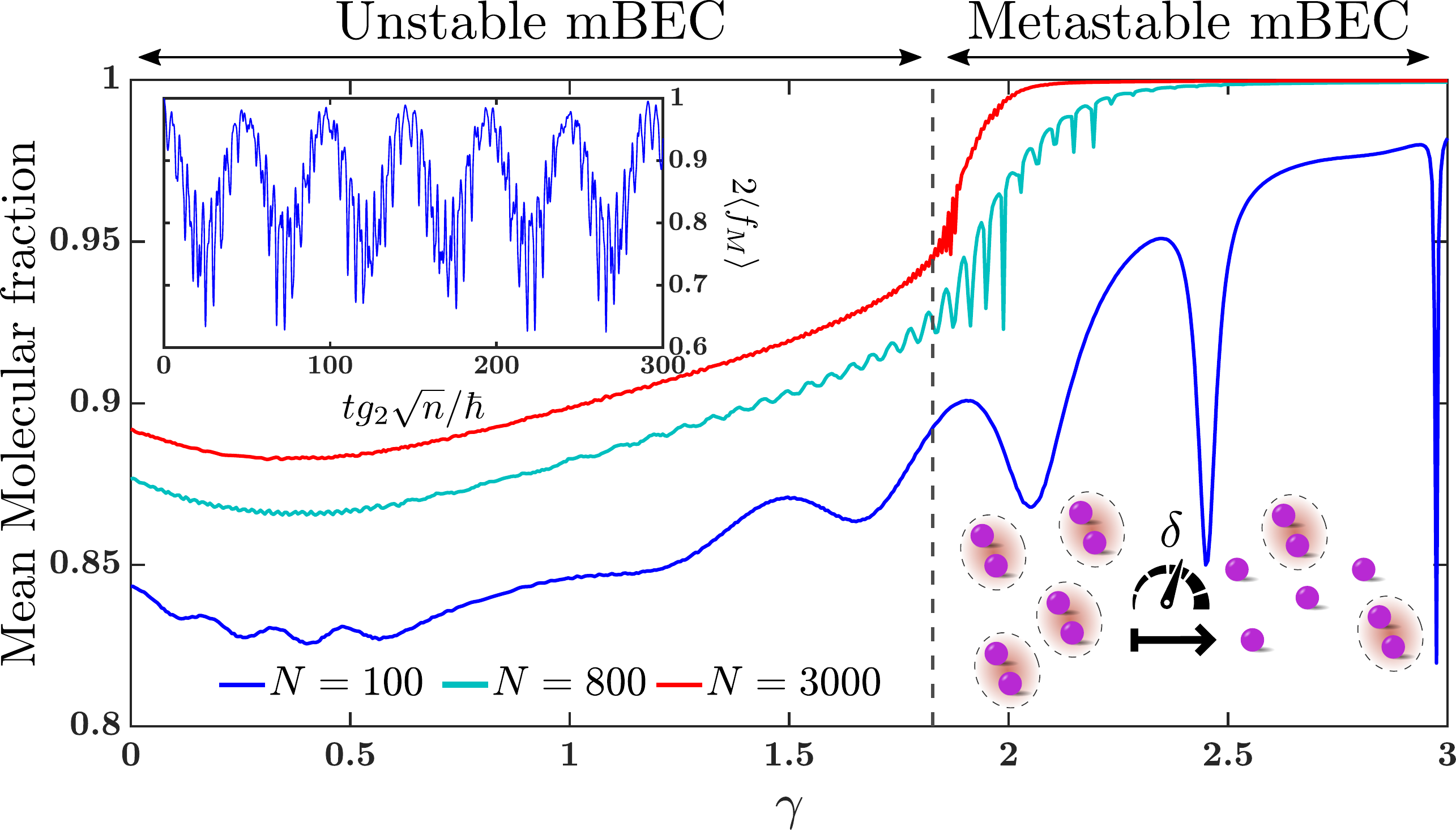}
\caption{
Molecular dissociation dynamics into a finite mixture of atoms (magenta balls) and molecules (balls in the dashed ellipses). 
A pure mBEC is prepared at large negative detunings, and subsequently $\delta$ is quenched to
$-\delta_c(\gamma)+1$ for different $\gamma$ and particle numbers (see legend). 
A quench for fixed $N$ entails $\gamma$, $g_3$ variation to attain $\delta_c(\gamma)$, while $g_2$ is kept constant. 
The time-averaged molecular fraction, $2\overline{\langle f_M \rangle}$, reveals  that the all-molecule state becomes metastable at large $\gamma$, due to the double-well structure.
The inset presents the dissociation dynamics of $N=100$ atoms at $\gamma=2.45$.
The vertical dashed line marks the point beyond which the mBEC becomes metastable.
}
\label{Fig:Dynamics}
\end{figure}

\paragraph*{\textit {Dynamical signatures.}} 

Molecular metastability in the presence of cTBR can be demonstrated by quench experiments. Namely, by preparing the system at large negative detuning so as to form a mBEC, quenching $\delta$ past $-\delta_c(\gamma)$, and monitoring the ensuing dissociation dynamics~\cite{Wang_stability_2024,moore2002bose}.
The resulting time-averaged molecular fractions $2\overline{\braket{f_M}}$ stemming from the two-mode Hamiltonian, where averaging is carried out over sufficiently long time to ensure convergence ($t=600\hbar/(g_2 \sqrt{n})$ here), are depicted in Fig.~\ref{Fig:Dynamics}. 
We note that incoherent three-body losses in an experiment may lead to damping of the coherent oscillations over time.
Up to $\gamma = 1.83$ [dashed line], the post-quench detuning lies in the unstable regime for the molecules [Fig.~\ref{Fig:Energy_contours}~(g)]. Hence, the time-averaged molecular fraction, $2\overline{\braket{f_M}}$ is less than unity. By contrast, for $\gamma > 1.83$ the all-molecules state is metastable [Fig.~\ref{Fig:Energy_contours}(g)], and $2\overline{\braket{f_M}}$ approaches unity. Since $1/N$ serves as an effective Planck constant, decreasing the number of particles elongates the tails of the quantum phase space distribution and enables tunneling to the ground state, i.e. loss of molecules. 
This process is responsible for the prominent dips in $2\overline{\langle f_M \rangle}$ curve, in the metastable mBEC regime for $N=100$, and is associated with the low-frequency dynamics of the molecular fraction [inset in Fig.~\ref{Fig:Dynamics}]. The low-frequency stems from the contribution of adjacent energy eigenstates close to avoided crossings in the energy spectrum of Eq.~\eqref{Eq:Hamiltonian}, see also SM~\cite{supp}.
These narrow avoided crossings originate from 
the presence of the double-well potential in the mean-field energy density.

\paragraph*{\textit {Summary \& Outlook.}} 

Coherent three-body recombination results in a first-order phase transition of atom-molecules to the molecular condensate, in contrast to the second-order associated with the Fano-Feshbach mechanism.
Such a structural change is rooted in the presence of a double-well in the mean-field energy density of the system.
The implications of such a double-well are two-fold.
First, an additional region of metastability opens up for the molecular condensate, and second, highly entangled macroscopic atom-molecule ensembles can be generated.

Our findings highlight coherent three-body processes as a versatile and experimentally accessible toolkit  for tuning the order of the ensuing quantum phase transitions and their entanglement in ultracold atom-molecule mixtures, inspiring forthcoming applications for quantum control of chemical reactions at ultralow temperatures. 
In this sense, developing a microscopic theory for cTBR is desirable. 
The predicted first-order transition, bistability, and metastable behavior await experimental observation which can be potentially achieved using magnetic ramp protocols across atom–molecule resonances. 
Accordingly, exploring dynamical criticality and classification to universality classes is another intriguing possibility.
Extending the present framework to account for finite-temperature effects and external confinement is of immense interest. Such a framework will reveal the impact of the competition between quantum and thermal fluctuations on the underlying chemical reactions and may reveal intriguing nonlinear dynamical regimes.

\paragraph*{\textit {Acknowledgments.}}

This work is financially supported by the Army Research Office (ARO) under Award number: W911NF-26-1-A043 (S.I.M.), the NSF under Grant No. PHY1511696 and PHY-2103542 (C.C.), the Air Force Office of Scientific Research under award
number FA9550-21-1-0447 (C.C.).
H.R.S. acknowledges support for ITAMP by the NSF.
G.A.B. and S.I.M. are grateful to J.C. Pelayo about extensive discussions on the Hartree-Fock-Bogoliubov framework.

\putbib[Refs]

\end{bibunit}

\clearpage

\begin{bibunit}[apsrev4-2]


\onecolumngrid
\setcounter{equation}{0}
\setcounter{figure}{0}
\setcounter{section}{0}
\makeatletter
\renewcommand{\theequation}{S\arabic{equation}}
\renewcommand{\thefigure}{S\arabic{figure}}
\renewcommand{\bibnumfmt}[1]{[S#1]}
\renewcommand{\citenumfont}[1]{S#1}
\renewcommand{\thesection}{\arabic{section}}
\setcounter{page}1
\def\thepage{S\arabic{page}}

\begin{center}
	{\Large\bfseries Supplementary Material: First-order phase transition in atom-molecule quantum degenerate mixtures with coherent three-body recombination \\ 
 }
\end{center}

\section{Atom-Molecule Hamiltonian} \label{Sec:Hamiltonian}

The 
Hamiltonian describing the atom-molecule mixture featuring the interplay of Feshbach coupling and cTBR takes the following form~\cite{Holland_formation_2001,Kokkelmans_ramsey_2002},
\begin{gather}
\hat{H} = \int d \boldsymbol{r} ~ \Bigg\{  \hat{\Psi}^{\dagger}_M(\boldsymbol{r}) \left[ -\frac{\hbar^2 \nabla^2}{4m}  + \Delta  \right] \hat{\Psi}_M(\boldsymbol{r}) + \hat{\Psi}^{\dagger}_A(\boldsymbol{r}) \left[ -\frac{\hbar^2 \nabla^2}{2m}  \right] \hat{\Psi}_A(\boldsymbol{r})  \nonumber \\  
 + \frac{g_{AA}}{2} \left[\hat{\Psi}^{\dagger}_A(\boldsymbol{r})\right]^2 \hat{\Psi}^2_A(\boldsymbol{r}) + \frac{g_{MM}}{2} \left[\hat{\Psi}^{\dagger}_M(\boldsymbol{r})\right]^2 \hat{\Psi}^2_M(\boldsymbol{r}) \nonumber \\
+ g_2 \left( \hat{\Psi}^{\dagger}_M(\boldsymbol{r}) \hat{\Psi}_A^2(\boldsymbol{r}) + \rm{h.c.}  \right) +   g_3 \left( \hat{\Psi}^{\dagger}_M(\boldsymbol{r})  \hat{\Psi}^{\dagger}_A(\boldsymbol{r}) \hat{\Psi}^3_A(\boldsymbol{r}) + \rm{h.c.} \right) \Bigg\}.  
\label{Eq:Hamiltonian_Supp_full}
\end{gather}
Here, $\hat{\Psi}_M(\boldsymbol{r})$ ($\hat{\Psi}_A(\boldsymbol{r})$) is the field operator annihilating a molecule (atom) at position $\boldsymbol{r}=(x,y,z)$.
Moreover, $\hbar$ is the reduced Planck constant, $m$ represents the atomic mass and $\text{h.c.}$ refers to the hermitian conjugate. 
Importantly, the parameter $\Delta$ is the detuning, $g_{AA}$ ($g_{MM}$) refers to the atom-atom (molecule-molecule) contact interaction strength, while $g_2$ ($g_3$) signifies the Feshbach (cTBR) coupling.  
The contact interactions are directly linked to the pertinent $s$-wave scattering lengths through the relations, $g_{AA} = 4\pi \hbar^2a_{AA}/m$, and $g_{MM}=2\pi \hbar^2 a_{MM}/m$. Note that atom-molecule collisions are mostly inelastic, leading to deep-dimer relaxation~\cite{Mukaiyama_Dissociation_2004,Yurovsky_atom_2000}, and therefore the atom-molecule scattering lengths remain experimentally  elusive. Typically, the recent ${}^{133}$Cs experiments are performed in box potentials where the atomic and molecular condensates assume a uniform density~\cite{Zhang_transition_2021}. 
Nevertheless, even external harmonic traps of frequency $\omega/(2\pi) \simeq 10 ~ \rm{Hz}$~\cite{zhang2023many} have a small effect as compared to the dominant Feshbach and cTBR couplings [see also Table~\ref{Tab:Energy_scales}]. Hence, in our analysis we neglect the external potentials.
Explicitly, the Feshbach coupling takes the following well-known expression~\cite{Duine_atom_2004}, $g_2 = \hbar \sqrt{2\pi a_{bg} \Delta B \Delta \mu /m}$, depending on the background atomic $s$-wave scattering length, $a_{bg}$, the width of the Feshbach resonance, $\Delta B$, and the difference of magnetic moments between a molecule and two atoms, $\Delta \mu$.
Notice here that the Feshbach coupling, $g_2$, is density independent and has units of $[E][V]^{1/2}$. On the other hand, the units of cTBR coupling are $[E][V]^{3/2}$. 

While no microscopic theory exists yet for the cTBR coupling, $g_3$, it may be linked to the three-body and four-body collision rates~\cite{Greene_review}. 
This is demonstrated by adiabatic elimination of the molecular field: Assuming a negligible and almost stationary molecular fraction in the Heisenberg equations of motion of the Hamiltonian of Eq.~\eqref{Eq:Hamiltonian_Supp}, the molecular field operator can be expressed in terms of the atomic operator, $\hat{\Psi}_A$. Substitution back to the pertinent Heisenberg equation, results in an effective equation for the atoms, 
\begin{equation}
i \hbar \frac{d}{dt} \hat{\Psi}_A = - \frac{\hbar^2 \nabla^2}{2m}\hat{\Psi}_A  + \left[ g_{AA} -  \frac{2 g_2^2}{\Delta} \right] \hat{\Psi}^{\dagger}_A \hat{\Psi}_A \hat{\Psi}_A
- \frac{6 g_2 g_3}{\Delta} \hat{\Psi}_A^\dagger \hat{\Psi}_A^\dagger \hat{\Psi}_A \hat{\Psi}_A \hat{\Psi}_A 
- \frac{3 g_3^2}{\Delta} (\hat{\Psi}_A^\dagger)^2 \hat{\Psi}_A \hat{\Psi}_A^\dagger \hat{\Psi}_A^3
- \frac{g_3^2}{\Delta} (\hat{\Psi}_A^\dagger)^3 \hat{\Psi}_A^4. 
\end{equation}
In the mean-field (large $N$ limit), field operators are replaced by $c$-numbers $\hat{\Psi}_A\rightarrow\Phi_A$, and the above equation reduces to
\begin{equation}
i \hbar \frac{d}{dt} \Phi_A = - \frac{\hbar^2 \nabla^2}{2m} \Phi_A +\left[ g_{AA} -
  \frac{2 g_2^2}{\Delta} \right]  n_A \Phi_A
- \frac{6 g_2 g_3}{\Delta} n_A^2 \Phi_A
- \frac{4 g_3^2}{\Delta} n_A^3 \Phi_A, 
\label{Eq:Effective_atoms}
\end{equation}
where $n_A = |\Phi_A|^2$. The second term models the two-body interaction arising due to Feshbach resonances~\cite{chin2010feshbach,Tommasini_Feshbach_1998}, while the next two refer to three- and four-body atomic interactions respectively, enhanced due to the presence of the molecules. In this sense, measurements of the three- and four-body collision rates may provide further insights on the cTBR coupling.

The various coupling mechanisms introduced above correspond to well distinctive energy scales. To explicate this argument, we consider the recently realized ${}^{133}$Cs atom-molecule mixture characterized by total peak density $n$. A clear separation of energy scales manifests, as demonstrated in Table~\ref{Tab:Energy_scales}, with the molecular detuning, Feshbach, and cTBR couplings being the most relevant contributions, see also the last section where all the terms are taken into account within a Hartree-Fock Bogoliubov approach.
As a consequence, here we focus on a minimal Hamiltonian capturing the basic features of atom-molecule mixtures, i.e.
\begin{equation}
\hat{H} = \int d \boldsymbol{r} ~ \Bigg\{ \Delta \hat{\Psi}^{\dagger}_M(\boldsymbol{r})  \hat{\Psi}_M(\boldsymbol{r})    
+ g_2 \left( \hat{\Psi}^{\dagger}_M(\boldsymbol{r}) \hat{\Psi}_A^2(\boldsymbol{r}) + \rm{h.c.}  \right) +   g_3 \left( \hat{\Psi}^{\dagger}_M(\boldsymbol{r})  \hat{\Psi}^{\dagger}_A(\boldsymbol{r}) \hat{\Psi}^3_A(\boldsymbol{r}) + \rm{h.c.} \right) \Bigg\}.  
\label{Eq:Hamiltonian_Supp}
\end{equation}

\begin{table}[t!] 
\setlength{\tabcolsep}{20pt}
\renewcommand{\arraystretch}{2}
\centering
\begin{tabular}{c|c|c|c|c|c|c}  \hline \hline
$E_b/h$ & $E_{AA}/h$ &  $E_{MM}/h$ & $E_F/h$ & $E_{\rm{cTBR}}/h$  & $E_T/h$ & $E_K/h$ \\ \hline
$2~\rm{kHz}$ & $239 ~ \rm{Hz}$ & $161~ \rm{Hz}$ & $1.2~ \rm{kHz}$ & $1.2 ~ \rm{kHz}$ & $10 ~ \rm{Hz}$ & $100 ~ \rm{Hz}$ \\ \hline \hline
\end{tabular}
\caption{
Characteristic energy scales for the molecular binding energy, $E_b \sim \abs{\Delta}$, atom-atom interaction, $E_{AA} \sim g_{AA}n$, molecule-molecule interactions, $E_{MM} \sim g_{MM}n$, Feshbach coupling, $E_F \sim g_2 \sqrt{n}$, cTBR coupling, $E_{\rm{cTBR}} \sim \gamma g_2 \sqrt{n}$, trapping energy, $E_T \sim \hbar \omega$, and kinetic energy, $E_K$, for ${}^{133}$Cs atom-molecule mixtures with peak density, $n$. The molecular binding energy depends directly on the detuning and in the range $\delta \in [-5,0]$, it holds that $E_b/h \in [4.2,0] ~ \rm{kHz}$. The kinetic energy is directly linked to non-condensed modes, and hence given their small fractions (see also last section), the kinetic energy is subdominant.
}
\label{Tab:Energy_scales}
\end{table}

\section{Atom-molecule mean-field energy with Feshbach coupling and cTBR} \label{Sec:Energy}

Within the two-mode approximation we assume that both atoms and molecules occupy a single box state, i.e. $\hat{\Psi}_A(\boldsymbol{r})= \frac{\hat{a}}{\sqrt{V}}$, and $\hat{\Psi}_M(\boldsymbol{r}) = \frac{\hat{b}}{\sqrt{V}}$, where $V$ is the box volume.
In this vein, the Hamiltonian of Eq.~\eqref{Eq:Hamiltonian_Supp} can be re-arranged as follows, 
\begin{equation}
    \hat{H} = \Delta \hat{b}^{\dagger} \hat{b} + \frac{g_2}{\sqrt{V}} \left(  \hat{b}^{\dagger} \hat{a}^2 + \text{h.c.}  \right) + \frac{g_3}{V^{3/2}} \left(  \hat{b}^{\dagger} \hat{a}^{\dagger} \hat{a}^3  + \text{h.c.}  \right),
    \label{Eq:Hamiltonian_two_mode_Supp}
\end{equation}
where $\hat{a}$ ($\hat{b}$) stands for the bosonic annihilation operator of an atom (molecule).

The mean-field description  of the many-body Hamiltonian of Eq.~(\ref{Eq:Hamiltonian_Supp}) is obtained in the limit of large $N$, by replacing the field operators with $c$-numbers, i.e. $\hat{\Psi}_{A,M}(\boldsymbol{r}) \to \Phi_{A,M}(\boldsymbol{r})$. 
Subsequently, the classical fields are expressed in the polar decomposition $\Phi_j(\boldsymbol{r}) = \sqrt{n_j} e^{i\theta_j}$, where $n_j$ ($\theta_j$) is the density (phase) of the $j$-species, $j=A,M$.
As such, it is possible to  retrieve the following energy density,
\begin{equation}
\mathcal{E} = \Delta n_M + 2 g_2 \sqrt{n_M} (n-2n_M) \cos(\varphi) + 2 g_3 \sqrt{n_M} (n-2n_M)^2 \cos(\varphi).\label{Eq:Energy_density_Supp}
\end{equation}
In this expression, $\varphi=2\theta_A-\theta_M$ is the phase difference between atoms and molecules whose densities $n_A$ and $n_M$ satisfy the  
constraint, $n_A+2n_M = n$, with $n$ being the total density.
To simplify the energy density, we employ the fraction $f_M=n_M/n$, and express $\mathcal{E}$ in units of $g_2 n^{3/2}$,
\begin{equation}
\frac{\mathcal{E}}{g_2 n^{3/2}} =    \delta f_M +2 \sqrt{f_M}(1-2f_M) \cos(\varphi)   + 2 \gamma \sqrt{f_M}(1-2f_M)^2 \cos(\varphi), 
\label{Eq:Energy_density_dim_Supp}
\end{equation}
with $\gamma=g_3n/g_2$.

To establish a connection between the two-mode Hamiltonian and the mean-field energy density, we express the former in units of $g_2 \sqrt{n}$, i.e.
\begin{equation}
   \frac{\hat{H}}{g_2 \sqrt{n}} =    \delta ~ \hat{b}^{\dagger} \hat{b} + \frac{1}{\sqrt{N}} \left(  \hat{b}^{\dagger} \hat{a}^2 + \text{h.c.}  \right) + \frac{\gamma}{N^{3/2}} \left(  \hat{b}^{\dagger} \hat{a}^{\dagger} \hat{a}^3  + \text{h.c.}  \right).
   \label{Eq:Hamiltonian_two_mode_Supp_dim}
\end{equation}
Without loss of generality, it is convenient to set the coefficient in front of the Feshbach conversion mechanism to unity, and therefore in our numerical treatment the following Hamiltonian is considered,
\begin{equation}
      \frac{\hat{H} \sqrt{V}}{g_2} =    \delta \sqrt{N} ~ \hat{b}^{\dagger} \hat{b} +  \left(  \hat{b}^{\dagger} \hat{a}^2 + \text{h.c.}  \right) + \frac{\gamma}{N} \left(  \hat{b}^{\dagger} \hat{a}^{\dagger} \hat{a}^3  + \text{h.c.}  \right).
      \label{Eq:Hamiltonian_two_mode_Supp_dim_2}
\end{equation}

\subsection{First-order transition with cTBR}  \label{Sec:Energy_TBR}

\begin{figure}[t!]
\centering
\includegraphics[width=1\columnwidth]{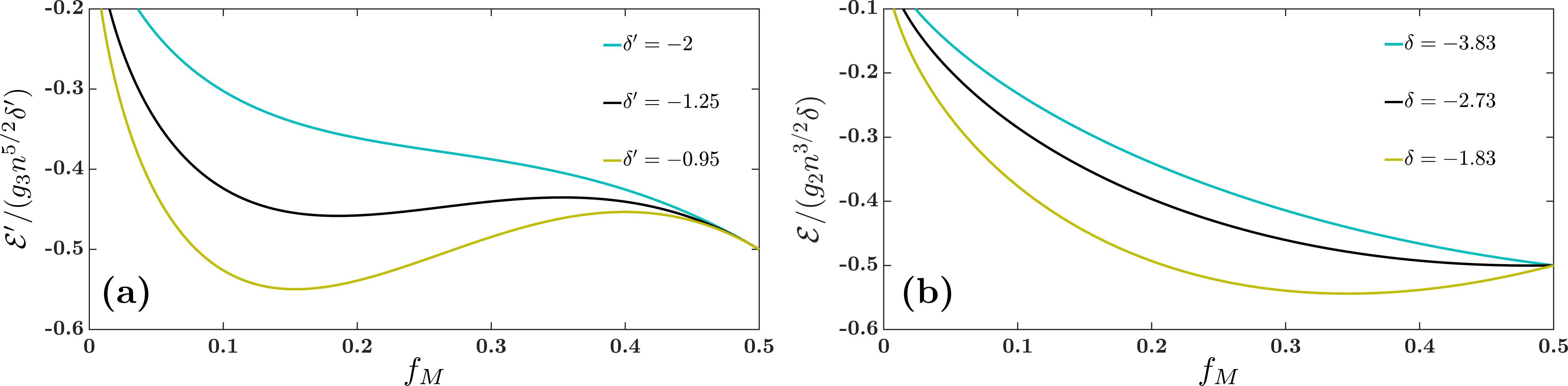}
\caption{Energy density profile at $\varphi=\pi$ associated to (a) cTBR [Eq.~\eqref{Eq:Energy_density_dim_TBR_Supp}] and (b) Feshbach coupling [Eq.~\eqref{Eq:Energy_density_dim_Supp} at $\gamma = 0$]. As it can be seen, in the presence of cTBR (panel (a)) for larger negative values of the detuning, e.g. $\delta=-2$, the energy density features a global minimum at $f_M =0.5$ where the molecular condensate is the ground state of the system. However, for decreasing $\delta$ the energy gradually deforms into a double-well structure possessing a global minimum at $f_M \approx 0.15$ (representing the ground state of the atom-molecule mixture) and a local one at $f_M=0.5$ (all molecule state). The local maximum at $f_M \approx 0.4$ refers to the saddle point. In contrast, when solely Feshbach coupling is present (panel (b)) a double-well structure is absent.   The energy density is normalized to its minimum, and is presented at different detunings (see legend). 
}
\label{Fig:Energy_density}
\end{figure}

To gain a solid understanding on the determination of the phase transition from the energy density, we first focus on the cTBR alone.
In the latter case the energy density reads,
\begin{equation}
    \frac{\mathcal{E}'}{g_3 n^{5/2}} = \delta' f_M +2\sqrt{f_M}(1-2f_M)^2 \cos(\varphi).
    \label{Eq:Energy_density_dim_TBR_Supp}
\end{equation}
The stationary points of the energy density surface, are defined by $\partial \mathcal{E}'/\partial f_M = \partial \mathcal{E}'/\partial\varphi=0$. Therefore, all fixed points must have $\varphi=0,\pi$ (note that for $f_M=0,0.5$, the atom-molecule phase $\varphi$ is ill defined due to the lack of molecules or atoms, respectively) with the global minimum, supporting the quantum ground state of the system, lying on the $\varphi=\pi$ line (simply because $\cos\pi <\cos 0^{\circ}$ and $2\sqrt{f_M}(1-2f_M)^2>0$). The mean-field energy density along this line, plotted  in Fig.~\ref{Fig:Energy_density},  should therefore reflect the physics of the phase transition.

For large negative detuning, the global minimum is the all-molecules state corresponding to $f_M=1/2$. It is not strictly a critical point of the energy surface because $\frac{\partial \mathcal{E}'}{\partial f_M} \Big|_{f_M=1/2}\neq 0$ but it minimizes the energy within the restricted $f_M\in[0,0.5]$ range (see e.g. $\delta'=-2$, $\delta'=-1.25$ in Fig.~\ref{Fig:Energy_density}(a)). Thus, to determine whether the $f_M=1/2$ point is a global minimum at a given fixed detuning $\delta'$, one simply needs to compare the energy of other fixed points to $\mathcal{E}'(f_M=0.5)/(g_3 n^{5/2})=\delta'/2$. For $\delta'>-1.51$, two such critical points, corresponding to the new minimum and the saddle point of the double-well structure, are found from $\frac{\partial \mathcal{E}'}{\partial f_M}=0$, see in particular $\delta' = -0.95$ in Fig.~\ref{Fig:Energy_density}(a). The first-order phase transition is obtained at $-\delta'_c$, where that minimum's energy drops below $ \mathcal{E}'(f_M=0.5)/(g_3n^{5/2})$ (i.e. below $\delta'/2$) and the ground state hops from one well ($f_M=1/2$) to the other ($f^{(c)}_M \simeq 0.15$). Thus, the critical detuning, $-\delta'_c$ and fraction, $f^{(c)}_M$ are set by the following two conditions, 
\begin{subequations}
    \begin{gather}
    \mathcal{E}' |_{f_M=1/2} = \mathcal{E}'|_{f_M=f^{(c)}_M,\varphi=\pi}, \\  \frac{\partial \mathcal{E}'}{\partial f_M} \Big |_{f_M=f^{(c)}_M,\varphi=\pi}=0,
\end{gather}
\label{Eq:Critical_TBR}
\end{subequations}
resulting in $\delta'_c = \frac{4\sqrt{2}}{3\sqrt{3}}$ and $f^{(c)}_M = \frac{1}{6}$.
Additionally, we identify the detuning at which the double-well structure emerges and disappears, $\mp \delta_d'$, by the presence of a stationary inflection point, namely by requiring that both $\frac{\partial \mathcal{E'}}{\partial f_M}$ and $\frac{\partial^2 \mathcal{E'}}{\partial f_M^2}$ vanish somewhere along either $\varphi=\pi$ or $\varphi=0$.
This demand is only satisfied at $\delta'=-\delta_d'$ at $\varphi=\pi$, and $\delta'=\delta_d'$ at $\varphi=0$, with $\delta'_d = \frac{8}{3} \sqrt{\frac{2}{5} \left( 4\sqrt{6} - 9  \right)} \simeq 1.51$.

\subsection{Phase transition with combined reaction mechanisms} \label{Sec:Phase_combined}

\begin{figure}[t!]
\centering
\includegraphics[width=0.6\textwidth]{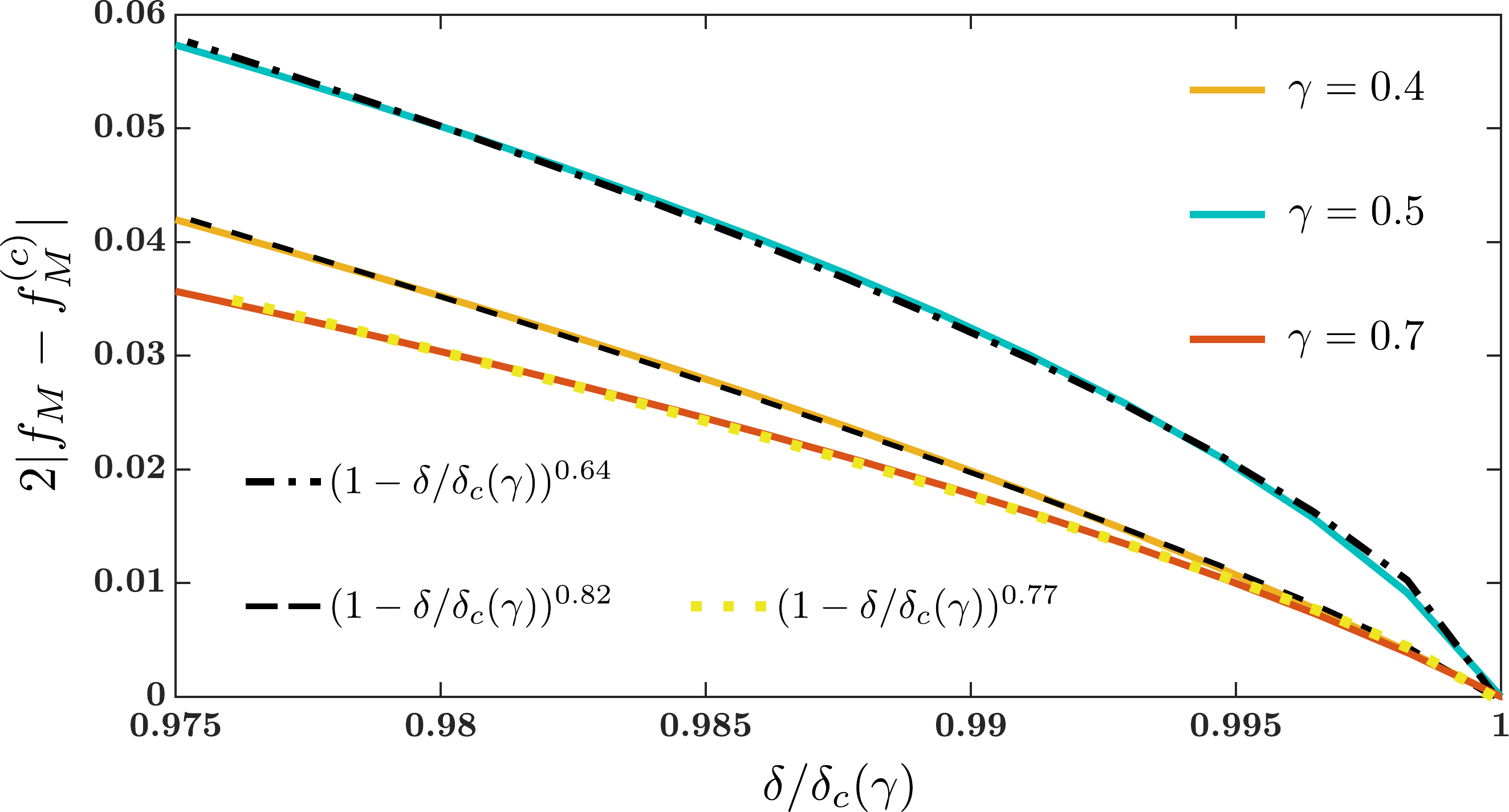}
\caption{
Scaling behavior of the molecular fractions, $2|f_M-f^{(c)}_M|$, in the vicinity of the quantum phase transition, $\delta_c(\gamma)$, for three representative $\gamma$ values (see legend) located slightly below ($\gamma=0.4$), above ($\gamma=0.7$) and at ($\gamma=0.5$) the tricritical point. The dashed, dotted and dash-dotted lines correspond to power-law fits, $(1-\delta/\delta_c(\gamma))^a$, within the interval $\delta \in [0.975,1]\delta_c(\gamma)$. The scaling exponents are shown in the legend. The exponents allude to a distinct scaling behavior at the tricritical point, $(\delta=-\delta_s,\gamma=0.5)$. 
}
\label{Fig:Criticality}
\end{figure}

Before delving into the energy landscape of both couplings, it is instructive to recap the second-order transition of the Feshbach mechanism alone [Eq.~\eqref{Eq:Energy_density_dim_Supp} at $\gamma = 0$].
Similarly to the case of cTBR, the all-molecules configuration ($f_M=1/2$) is a global minimum for small $\delta$ [Fig.~\ref{Fig:Energy_density}(b) at $\delta = -3.83$]. 
However, in contrast to cTBR [Fig.~\ref{Fig:Energy_density}(a)] the phase transition is continuous and does not involve any bifurcations: As $\delta$ is increased, the gradient $\frac{\partial \mathcal{E}}{\partial f_M} \Big|_{f_M=1/2}$ decreases until at $\delta = -\delta_s = -2\sqrt{2}$ becomes zero and the minimum shifts continuously to the critical point $f_M \lesssim 1/2$  for which $\frac{\partial \mathcal{E}}{\partial f_M}=0$ [see Fig.~\ref{Fig:Energy_density}(b) at $\delta = -2.73$]. As $\delta$ is further increased, this critical point corresponds to smaller and smaller molecular fractions (see e.g. $\delta = -1.83$ in Fig.~\ref{Fig:Energy_density}(b)). 
As a further confirmation of the absence of a double-well, no saddle points exist: the system of equations $\frac{\partial \mathcal{E}}{\partial f_M}=0$ and $\frac{\partial^2 \mathcal{E}}{\partial f_M^2}=0$ admits no solution at $\gamma = 0$.

Since the double-well potential in the energy density underlies the observation of a first-order phase transition, we infer its existence in the presence of both Feshbach and cTBR couplings, by identifying the onset and demise of the saddle point. A solution for the pertinent equations $\frac{\partial \mathcal{E}}{\partial f_M}=0$ and $\frac{\partial^2 \mathcal{E}}{\partial f_M^2}=0$, is admitted only for $\gamma \geq 0.5$. Explicitly, the double-well occurs within the range of detunings, $\delta \in [-\delta_d(\gamma),-2\sqrt{2}]\cup[2\sqrt{2},\delta_d(\gamma)]$, where
\begin{equation}
\delta_d(\gamma) = \frac {  \left(12\gamma+6 \right) \sqrt{ 32  \gamma  (\gamma + 1) + 3}  - 16\sqrt {3}  \left( \gamma^2 +\gamma-\frac{3}{8} \right)  } {3\sqrt {5} \sqrt{ \gamma \sqrt {96  \gamma  (\gamma + 1) + 
9 } + 3 \gamma + 6\gamma^2}}, \quad ~~\textrm{for}~~ \gamma \geq 0.5.
\label{Eq:Potential_existence_Supp}
\end{equation}

Having identified the $\gamma$ interval where a first-order phase transition takes place, we next proceed to determine the critical detuning, $-\delta_c(\gamma)$, where such a transition occurs.
Similarly to the pure cTBR case [see Eqs.~\eqref{Eq:Critical_TBR}], we require that when the double-well exists, the energy density of the all-molecules configuration ($f_M=0.5$) equals the energy of the other minimum at $f^{(c)}_M<0.5$, resulting in,
\begin{subequations}
    \begin{gather}
    \delta_c(\gamma) =  \frac{4 \sqrt{2} (1+\gamma)^{3/2}}{3 \sqrt{3\gamma} },  \label{Eq:Critical_detuning_Supp} \\
    f^{(c)}_M = \frac{1 + \gamma}{6 \gamma},
    \label{Eq:Critical_pop_Supp} 
\end{gather}
\end{subequations}
for $\gamma \geq 0.5$.
Setting $\gamma = 0.5$ we have $\delta_c(0.5) =\delta_s = 2\sqrt{2}$ and $f^{(c)}_M=1/2$, i.e. one retrieves the critical values of the second-order phase transition occurring at $\gamma = 0$.
In fact, $-\delta_s$ is the detuning characterizing the second-order transition in the entire $\gamma \in [0,0.5)$ interval.
This is directly inferred from the condition $\frac{\partial \mathcal{E}}{\partial f_M} \Big |_{f_M=1/2}=0$, i.e. at the threshold of appearance of a finite atom-molecule mixture [see also Fig.~\ref{Fig:Energy_density}(b) at $\delta = -2.73$].

The $(\delta=-\delta_s,\gamma=0.5)$ point is a tricritical point since it lies right at the junction of the first- and second-order transition boundary, see also Fig.~1(c) in the main text.
To further understand the emergence of tricriticality, one needs to analyze the scaling exponents that appear near the phase transitions.
Specifically, the scaled molecular fraction $2|f_M-f^{(c)}_M|$ displays a power-law behavior in the vicinity of the critical detuning, $\delta_c(\gamma)$, $ \propto (1-\delta/\delta_c(\gamma))^a$, as demonstrated in Fig.~\ref{Fig:Criticality}. The different exponent ($a=0.64$) at $\gamma=0.5$, compared to the value obtained for other coupling ratios ($a=0.82$, $a=0.77$), indicates that the tricritical $(\delta=-\delta_s,\gamma = 0.5)$ point is characterized by a distinct scaling behavior. 
Nevertheless, a more rigorous analysis of the critical behavior, its associated scaling exponents and possible universality classes is required, which we defer to future studies since it is a subject worth pursuing on its own right.

\begin{figure}[t!]
\centering
\includegraphics[width=1\columnwidth]{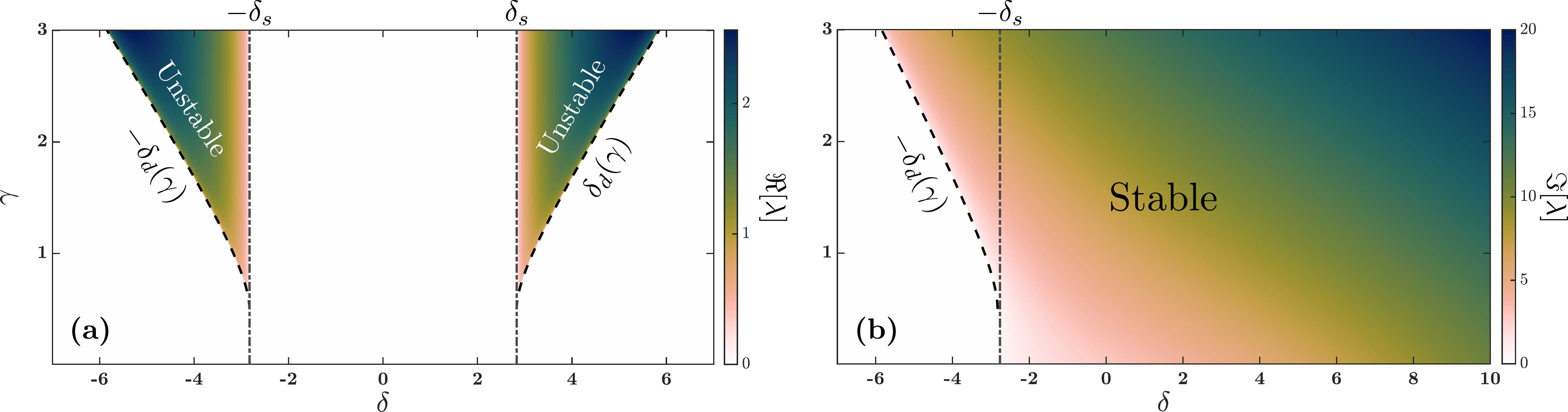}
\caption{Eigenvalues of the Jacobian stemming from the linearized Hamiltonian system around (a) the saddle points referring to the double-well potential and (b) the minimum associated with the atom-molecule mixture for $\delta \geq -\delta_d(\gamma)$, with $\gamma \geq 0.5$ and for $\delta \geq - \delta_s$, where $\gamma<0.5$. 
The dashed and dash-dotted lines mark the regions of existence of these stationary points (see legends). 
The finite atom-molecule mixture is a stable point, as inferred from the purely imaginary eigenvalues [panel (b)], in contrast to the saddle points which are unstable [panel (a)].}
\label{Fig:Eigenvalues}
\end{figure}

\section{Stability analysis of the atom-molecule system} \label{Sec:Stability}

The energy density given by  Eq.~\eqref{Eq:Energy_density_dim_Supp} [Eq.~(2) in the main text] can be viewed as the rescaled Hamiltonian $H= \frac{\mathcal{E} V}{g_2 \sqrt{n}}$, governing the dynamics of the atom-molecule mixture. 
The state of the latter is characterized by the dynamical variables $f_M,\varphi$, which evolve according to the Hamilton's equations, $\dot{f}_M= \frac{\partial H}{\partial \varphi}$ and $\dot{\varphi} = - \frac{\partial H}{\partial f_M}$.
All relevant fixed points are determined by solving the set of equations $\dot{f}_M=0$, $\dot{\varphi}=0$, and they are labeled as $(f^{(0)}_M,\varphi^{(0)})$.
The only exception is $f^{(0)}_M=1/2$.
Even if $\dot{\varphi} \neq 0$, the $f^{(0)}_M=1/2$ is still a stationary point since the relative phase $\varphi$ is ill-defined in the absence of atoms.

To assess the stability of the stationary points, we first write the Hamilton's equations in a compact matrix form,
\begin{equation}
\dot{\boldsymbol{z}} = \boldsymbol{J} \cdot \boldsymbol{\nabla} H(z), \quad \text{where} \quad \boldsymbol{z}= (f_M,~\varphi)^{\top},  \quad \boldsymbol{J} = \begin{pmatrix}
 0 & 1 \\
-1 & 0
\end{pmatrix}, \quad \boldsymbol{\nabla}H(z) = \left(  \frac{\partial H}{\partial f_M},~\frac{\partial H}{\partial \varphi}  \right)^{\top}.
\label{Eq:Hamilton_Supp}
\end{equation}
In particular, we are interested in small perturbations around the equilibrium points $\boldsymbol{z}^{(0)} = (f^{(0)}_M,~\varphi^{(0)})^{\top}$, with $\boldsymbol{\delta z} = \boldsymbol{z}- \boldsymbol{z}^{(0)}$.
The linearized Hamilton's equations then read  
\begin{equation}
\boldsymbol{\delta \dot{z}} = \boldsymbol{J} \cdot \boldsymbol{\nabla^2 H}(\boldsymbol{z}^{(0)}) \cdot \boldsymbol{\delta z},~~~\textrm{with}~~~ \boldsymbol{\nabla^2 H}(\boldsymbol{z}^{(0)}) = \begin{pmatrix}
\frac{\partial^2 H}{\partial f_M^2} & \frac{\partial^2 H}{\partial f_M \partial \varphi} \\
\frac{\partial^2 H}{\partial \varphi \partial f_M} & \frac{\partial^2H}{\partial \varphi^2}
\end{pmatrix}  \Bigg \lvert_{\boldsymbol{z}^{(0)}}.
\label{Eq:Linearization_Supp}
\end{equation}
The matrix $\boldsymbol{\nabla^2 H}(\boldsymbol{z}^{(0)})$ is the Hessian matrix associated with the Hamiltonian at $\boldsymbol{z}^{(0)}$, whereas the matrix product $\boldsymbol{J} \cdot \boldsymbol{\nabla^2 H}(\boldsymbol{z}^{(0)})$ is the Jacobian~\cite{strogatz2024nonlinear}. 
The general solution of Eq.~\eqref{Eq:Linearization_Supp} is $\boldsymbol{\delta z}(t)=\sum_{i=1}^2 c_i e^{\lambda_i t} \boldsymbol{u}_i$, where $\lambda_i$ ($\boldsymbol{u}_i$) are the eigenvalues (eigenvectors) of the Jacobian, while $c_i$ are the projection coefficients of the initial condition, $\boldsymbol{\delta z}(0)$, to the eigenvector basis. 
Thus, linear stability is ensured if both eigenvalues, $\lambda$, of the Jacobian are purely imaginary. 
The relevant characteristic polynomial for the eigenvalues is $\lambda^2 + \det\left[\boldsymbol{\nabla^2H}(\boldsymbol{z}^{(0)})\right]=0 \Longrightarrow \lambda = \pm \sqrt{- \det \left[ \boldsymbol{\nabla^2 H}(\boldsymbol{z}^{(0)})  \right]}$.
Therefore, if the determinant of the Hessian matrix is positive, it means that $\boldsymbol{z}^{(0)}$ is an elliptic (i.e. stable) point, whereas if $\det \left[  \boldsymbol{\nabla^2 H}(\boldsymbol{z}^{(0)})  \right] <0$, then $\boldsymbol{z}^{(0)}$ is a hyperbolic (namely unstable) point.
The determinant of the Hessian is easily calculated and takes the following general expression,
\begin{gather}
  \det \left[ \boldsymbol{\nabla^2 H}(\boldsymbol{z}^{(0)})  \right] = \nonumber \\  \frac {\left [(4  f^{(0)}_M  (5  f^{(0)}_M - 1) + 1)  (\gamma - 2  \gamma  f^{(0)}_M)^2 + 
       \gamma  (2 - 4  f^{(0)}_M  (2  f^{(0)}_M  (6  f^{(0)}_M - 5) + 3)) + 4  f^{(0)}_M  (3  f^{(0)}_M - 1) + 
       1 \right]\cos  (2  \varphi^{(0)})} {f^{(0)}_M} \nonumber \\ - 
 8  \gamma  (2  f^{(0)}_M - 1)  (2  \gamma  (f^{(0)}_M  (10  f^{(0)}_M - 7) + 1) - 12  f^{(0)}_M + 3) - 24  f^{(0)}_M + 
 8.
  \label{Eq:Hessian-Supp}
\end{gather}

\begin{figure}[t!]
\centering
\includegraphics[width=1\columnwidth]{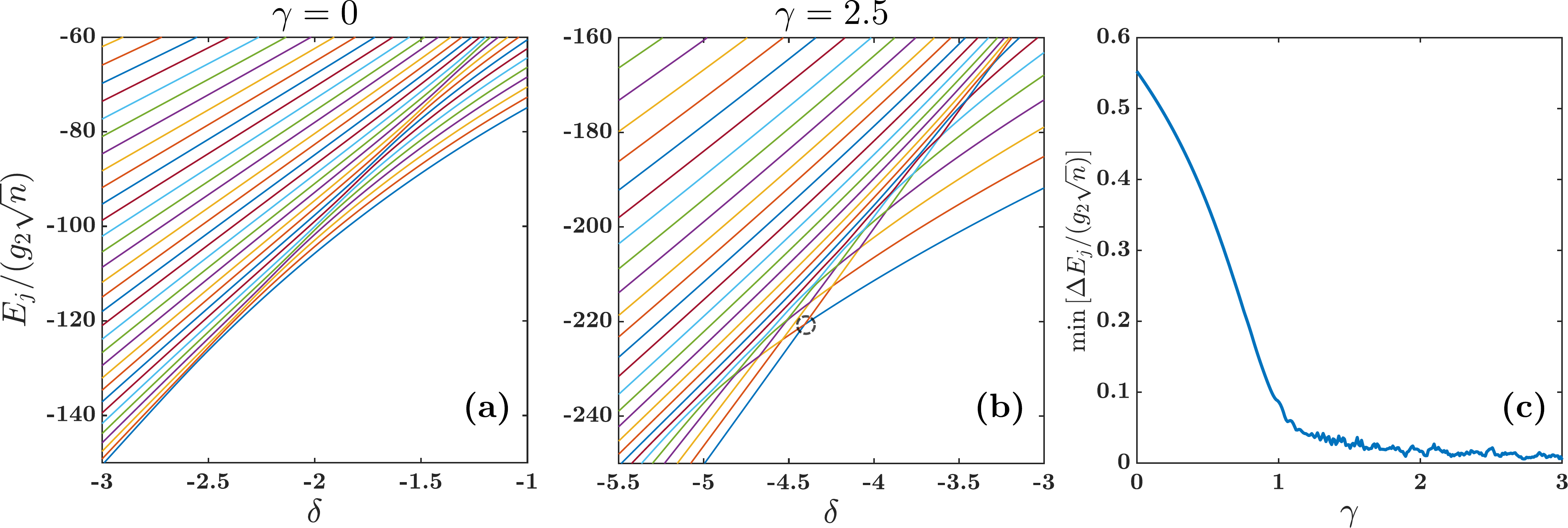}
\caption{
Energy eigenvalues, $E_j,~j=0,\ldots,29$, of the two-mode Hamiltonian [Eq.~(1) in the main text] for $N=100$ atoms at (a) $\gamma = 0$ and (b) $\gamma = 2.5$. 
The different colors represent distinct eigenvalues $j$.
The energy spectra are presented in the vicinity of the second-order (first-order) phase transition, i.e. $\delta = -\delta_s$ ($\delta = -4.46$). 
The dashed circle in panel (b) marks the position of a narrow avoided crossing. (c) Minimum of the energy difference between adjacent energy eigenvalues, $\Delta E_j = E_{j+1} - E_j$, over the detuning $\delta$. The avoided crossings become more narrow as $\gamma$ increases.}
    \label{Fig:Energy_spectrum}
\end{figure}

To visualize the stability of the atom-molecule mixture,  
we present the eigenvalues of the Jacobian for two stationary points across the $\delta$-$\gamma$ plane in Fig.~\ref{Fig:Eigenvalues}.
The saddle points emanating from the double-well potential [see also Fig.~2(b), (c) in the main text] are unstable, since the eigenvalues are purely real [Fig.~\ref{Fig:Eigenvalues}(a)].
Note that these points exist only in the regimes defined by the intersection of $\mp \delta_s$ and $\mp \delta_d(\gamma)$ [dash-dotted and dashed lines in Fig.~\ref{Fig:Eigenvalues}(a) respectively], since they are inherently linked to the double-well structure.
On the other hand, the minimum related to the presence of the atomic component [see also Fig.~2(b)-(f) in the main text] is stable, as evidenced by the purely imaginary eigenvalues of the Jacobian [Fig.~\ref{Fig:Eigenvalues}(b)].

\section{Von Neumann entropy for atoms and molecules} \label{Sec:Entropy}

The ground state, $\ket{\psi_0}$, describing the atom-molecule mixture is written as a linear superposition of number states,
\begin{equation}
    \ket{\psi_0} = \sum_{n_M=0}^{N/2} c_{n_M} \ket{n_M;N-2n_M}.
    \label{Eq:Superposition_Supp}
\end{equation}
The $c_{n_M}$ coefficients are the eigenvectors stemming from the diagonalization of the two-mode Hamiltonian.
Another way of representing the state of the ensemble is through the pure density matrix~\cite{nielsen2010quantum},
\begin{equation}
    \hat{\rho} = \ket{\psi_0} \bra{\psi_0}  = \sum_{n_M, n_M'} c_{n_M} c^*_{n_M'}  \ket{n_M; N-2n_M} \bra{n_M'; N-2n_M'}.
    \label{Eq:Density_matrix_Supp}
\end{equation}
The atom-molecule mixture is a bipartite system, and therefore to assess the degree of entanglement between the two constituents we employ the von Neumann entropy of the reduced density matrix of either atoms ($A$) or molecules ($M$)~\cite{Preskill_notes}.
Following the same prescription as in Hines \textit{et al.}~\cite{Hines_entanglement_2003} we trace over the atom degrees-of-freedom to obtain the molecular density matrix 
\begin{equation}
\hat{\rho}_M = \text{Tr}_A(\hat{\rho}  ) = \sum_{N_a=0}^N \sum_{n_M,n_M'=0}^{N/2} c_{n_M} c^*_{n_M'}  \braket{N_a | n_M;N-2n_M} \braket{n_M';N-2n_M' | N_a} = \sum_{n_M=0}^{N/2} \abs{c_{n_M}}^2 \ket{n_M} \bra{n_M}.
\label{Eq:Reduced_density_Supp}
\end{equation}
The above expression of the molecular density matrix is directly in a diagonal form, and the von Neumann entropy, $S$, is easily calculated 
\begin{equation}
S = - \text{Tr} \left[  \hat{\rho}_M \log_2(\hat{\rho}_M)   \right]  = - \sum_{n_M=0}^{N/2} \abs{c_{n_M}}^2 \log_2 \left(  \abs{c_{n_M}}^2  \right).
\label{Eq:Entropy_Supp}
\end{equation}
In particular, a product state of atoms and molecules, $\ket{\psi_0}= \ket{n_M';N-2n_M'}$, is characterized by $S=0$. We also denote the product state as a definite atom-molecule mixture, i.e. a single number state is required to describe the system.
In contrast, a maximally entangled atom-molecule setting occurs when all number states are populated with the same probability, i.e. $\abs{c_{n_M}}^2 = \frac{1}{1+N/2}$. In this case the system consists of a superposition of all possible atom-molecule states, characterized by the maximum entropy, $S=\log_2(1+N/2)$.
Even though a high degree of entanglement between atoms and molecules occurs near the phase transitions (irrespective of their order), such maximally entangled states are never attained.
It is finally worth mentioning that the maximal entropy is larger in the vicinity of the first-order phase transition as compared to the second-order one.

\section{Energy spectra of the two-mode Hamiltonian}
\label{Sec:Energy_spectra}

The presence of a double well potential in the mean-field energy affects the energy spectrum of the two-mode Hamiltonian as well [Eq.~(1) in the main text].
This is evident when inspecting the energy eigenvalues, $E_j,~j=0,\ldots,N/2$, in the parametric regime where a second- and first-order phase transition occur, see e.g. the energy spectra for $N=100$ in Fig.~\ref{Fig:Energy_spectrum}(a), (b).
The crucial difference is that avoided crossings become narrower in the case of a first-order transition (namely in the presence of cTBR), see the dashed circle in Fig.~\ref{Fig:Energy_spectrum}(b).
In fact, the energy eigenstates in the vicinity of the sharp avoided crossings are related to doublet states in the double well potential in the mean-field energy [see also Fig.~\ref{Fig:Energy_density}(a)].
This is further corroborated by inspecting the minimum energy gap between adjacent energy eigenvalues with respect to $\gamma$ in Fig.~\ref{Fig:Energy_spectrum}(c). The sharper avoided crossings for larger $\gamma$ can be attributed to the increase of the double well barrier height in the mean-field energy density [Eq.~\eqref{Eq:Energy_density_dim_Supp}].
Specifically, during the molecular dissociation dynamics tunneling occurs within these eigenstate doublets in the vicinity of their sharp avoided crossings. This process is subsequently imprinted on the prominent dips in the time-averaged molecular fraction presented in Fig.~4 of the main text.
However, in the case of larger atom numbers the aforementioned avoided crossings become narrower and hence tunneling is suppressed.

\begin{figure}[t!]
    \centering
    \includegraphics[width=1\columnwidth]{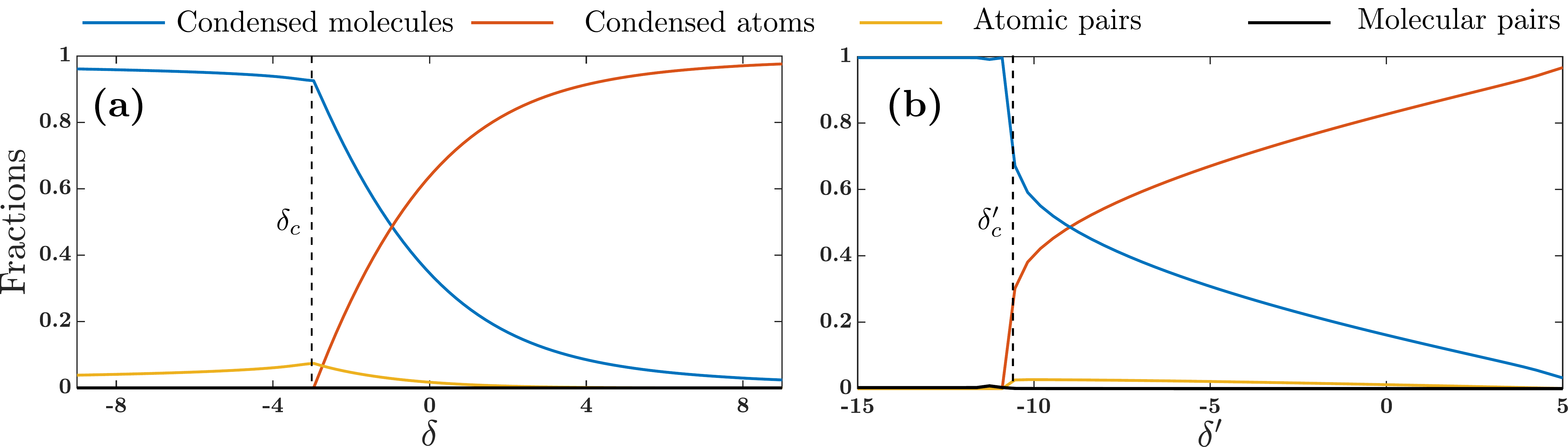}
    \caption{
    Phase diagrams displaying the population fractions stemming from the Hartree-Fock-Bogoliubov framework in the case of purely (a) Feshbach and (b) cTBR reaction mechanisms. The fractions of non-condensed atomic and molecular pairs (see legend) indicate that quantum depletion remains low in both cases. Despite the presence of quantum fluctuations, both second- [panel (a)] and first-order [panel (b)] phase transitions persist in the atom-molecule mixtures. The respective critical points, $\delta_c$ and $\delta_c'$, are marked by the vertical dashed lines.
    }
    \label{Fig:HFB}
\end{figure}

\section{Hartree-Fock-Bogoliubov framework}
\label{Sec:HFB}

As argued in the first Section above, a separation of energy scales associated with the different coupling terms emerges. 
This indicates that the minimal Hamiltonian [Eq.~\eqref{Eq:Hamiltonian_Supp}] is adequate for capturing  the most important features of the considered atom-molecule mixtures. In this section we further corroborate the above argument, and demonstrate that the impact of the additional coupling terms is small. This is achieved by treating the extended Hamiltonian within the Hartree-Fock-Bogoliubov framework~\cite{Yukalov_gapless_2006}, which is capable of capturing quantum fluctuations in a self-consistent way.
Focusing on homogeneous atom-molecule mixtures trapped within a box of volume $V$, the atomic and molecular field operators are expanded in terms of bosonic momentum modes, $\hat{a}_{\boldsymbol{k}}$, $\hat{b}_{\boldsymbol{k}}$, i.e. $\hat{\Psi}_A(\boldsymbol{r}) = \sum_{\boldsymbol{k}} e^{i \boldsymbol{k} \cdot \boldsymbol{r}} \hat{a}_{\boldsymbol{k}}/\sqrt{V}$, $\hat{\Psi}_M(\boldsymbol{r}) = \sum_{\boldsymbol{k}} e^{i \boldsymbol{k} \cdot \boldsymbol{r}} \hat{b}_{\boldsymbol{k}}/\sqrt{V}$. 
Upon employing the above field operator expansions, the many-body Hamiltonian takes 
the following form,

\begin{align}
\hat{H} = &\sum_{\boldsymbol{k}} \left[  \left(\frac{\hbar^2\boldsymbol{k}^2}{2m} -\mu \right)  a^{\dagger}_{\boldsymbol{k}} a_{\boldsymbol{k}} 
+  \left(\frac{\hbar^2\boldsymbol{k}^2}{4m} -2\mu \right)  b^{\dagger}_{\boldsymbol{k}} b_{\boldsymbol{k}}  \right]  + \frac{g_{AA}}{2V}\sum_{\boldsymbol{k}_1,\boldsymbol{k}_2,\boldsymbol{k}_3}  a^{\dagger}_{\boldsymbol{k}_1}a^\dagger_{ \boldsymbol{k}_2}a_{\boldsymbol{k}_3}a_{\boldsymbol{k}_1+\boldsymbol{k}_2-\boldsymbol{k}_3}+ \frac{g_{MM}}{2V}\sum_{\boldsymbol{k}_1,\boldsymbol{k}_2,\boldsymbol{k}_3}  b^{\dagger}_{\boldsymbol{k}_1}b^{\dagger}_{ \boldsymbol{k}_2}b_{\boldsymbol{k}_3}b_{\boldsymbol{k}_1+\boldsymbol{k}_2-\boldsymbol{k}_3} \nonumber \\
+& \frac{g_2}{\sqrt{V}}\sum_{\boldsymbol{k}_1,\boldsymbol{k}_2}\left(a^{\dagger}_{\boldsymbol{k}_1}a^{\dagger}_{\boldsymbol{k}_2}b_{\boldsymbol{k}_1+\boldsymbol{k}_2}+\rm{h.c.}\right)
+ \frac{g_3}{V^{3/2}}\sum_{\boldsymbol{k}_1,\boldsymbol{k}_2,\boldsymbol{k}_3,\boldsymbol{k}_4} \left(a^{\dagger}_{\boldsymbol{k}_1}a^{\dagger}_{\boldsymbol{k}_2}a^{\dagger}_{\boldsymbol{k}_3}b_{\boldsymbol{k}_4}a_{\boldsymbol{k}_1+\boldsymbol{k}_2+\boldsymbol{k}_3-\boldsymbol{k}_4} + \rm{h.c.} \right),
\end{align}
where $\mu$ is the chemical potential of the entire mixture.
Following the same prescription as in the recent studies of ultracold ${}^{133}$Cs atom-molecule mixtures~\cite{Wang_stability_2024,Wang_tunable_2025}, 
a suitable variational ansatz is employed in order to derive the equations of motion for the condensed ($\boldsymbol{k}=0$) and non-condensed ($\boldsymbol{k} \neq 0$) atoms and molecules, which are  subsequently solved numerically. Details on the construction and ingredients of the Hartree-Fock Bogoliubov approach in the presence of cTBR will be reported in a forthcoming publication~\cite{Bougas_unpublished}.

It turns out that atoms and molecules remain mostly condensed, while only a small population fraction ends up in the non-condensed modes, see Fig.~\ref{Fig:HFB}.
Importantly, a second-order (first-order) transition persists in the case of purely Feshbach (cTBR) coupling, as evidenced in Fig.~\ref{Fig:HFB}(a) (Fig.~\ref{Fig:HFB}(b)).
As it can be seen, agreement occurs between the Hartree-Fock-Bogoliubov [Fig.~\ref{Fig:HFB}(a)] and the two-mode approach [Fig.~1(a) in the main text] predictions on the second-order phase transition induced by Feshbach coupling. However, complete molecular conversion past the critical point is hindered in the Hartree-Fock-Bogoliubov due to the presence of atomic pairs.
On the other hand, the nature of the first-order transition remains intact, while its characteristics, such as the 
atom-molecule populations at the critical point, and the associated detuning, $\delta_c'$, seem to be slightly modified in the presence of quantum fluctuations that are related to the miscroscopic origin of the cTBR [Eq.~\eqref{Eq:Effective_atoms}].
As mentioned in the first Section, a treatment of the latter remains still elusive.

We emphasize that the phase diagrams presented in Fig.~\ref{Fig:HFB} take into account not only the quantum depletion, but also kinetic terms as well as atom-atom and molecule-molecule interactions.
The latter have only a small effect on the phase diagrams as can be deduced by systematically including these contributions one-by-one in the calculation of the ground state phase-diagrams (not shown). Additionally, a direct comparison of Fig.~\ref{Fig:HFB}(a), (b) and Fig.~1(a), (b) reveals that all these contributions do not affect the nature of the transitions but rather slightly shift their location. 
Finally, note that the Hartree-Fock-Bogoliubov formulation is suitable for addressing finite temperature effects by populating the non-condensed modes according to a Bose-Einstein distribution~\cite{Yukalov_gapless_2006}.
However, such effects become important only at temperatures far above the quantum degenerate regimes probed in recent experiments~\cite{zhang2023many,Wang_stability_2024} being the focus of our current study. 
Certainly, the impact of finite temperature effects close to or above the condensation temperature on the ensuing phases and their critical behavior is an intriguing direction on its own which will be explored in future studies.

\putbib[Refs]

\end{bibunit}

\end{document}